\documentclass[intlimits,twoside,a4paper]{article}

\usepackage{amsmath,amssymb}
\usepackage{graphicx}

\usepackage[eqsecnum]{cmpj2e}

\DeclareMathOperator{\Tr}{Tr}
\DeclareMathOperator{\Img}{Im}
\DeclareMathOperator{\Real}{Re}
\DeclareMathOperator{\Ind}{Ind}

\title[F-electron spectral function of the Falicov-Kimball model \ldots]%
{F-electron spectral function of the Falicov-Kimball model
and the Wiener-Hopf sum equation approach }
\author{A. M. Shvaika\refaddr{label1} and
        J. K. Freericks\refaddr{label2}}
\addresses{
\addr{label1} Institute for Condensed Matter Physics of the National Academy of Sciences of Ukraine, 1 Svientsitskii Str., 79011 Lviv, Ukraine
\addr{label2} Department of Physics, Georgetown University, 37th and O Sts. NW, Washington, DC 20057, USA
}
\begin{document}

\maketitle

\begin{abstract}
We derive an alternative representation for the $f$-electron spectral function
of the Falicov-Kimball model from the original one proposed by Brandt and
Urbanek. In the new representation, all calculations are restricted to the real
time axis, allowing us to go to arbitrarily low temperatures. The general 
formula for the retarded Green's function involves two determinants of
continuous matrix operators that have the Toeplitz form.  By employing the
Wiener-Hopf sum equation approach and Szeg\"o's theorem, we can derive exact analytic
formulas for the large-time limits of the Green's function; we illustrate this for cases when the logarithm of
characteristic function (which defines the continuous Toeplitz matrix) does and does not wind around the origin. We show how
accurate these asymptotic formulas are to the exact solutions found from
extrapolating matrix calculations to the zero discretization size limit.
\pacs 71.10.--w, 71.27.+a, 71.30.+h, 02.30.Rz
\end{abstract}

\section{Introduction}

The Falicov-Kimball model~\cite{falicov_kimball} has often been studied
as one of the simplest models of strong electron correlations.  Indeed,
the exact solution in the limit of large dimensions displays charge
density wave order, the Mott metal-insulator transition, and phase 
separation~\cite{freericks_review}.  It is particularly relevant to present a paper
on this model in an issue dedicated to Prof.~Stasyuk, as he has worked
on this model and closely related ones throughout his career; this began with his first article on the Falicov-Kimball model~\cite{stasyuk:FKM1} while his most relevant work to this contribution involves approximate solutions for the $f$-electron spectral functions~\cite{stasyuk:FKM_f1,stasyuk:FKM_f2,stasyuk:FKM_f3}. 

Our focus in this work is on the spectral function of the $f$-electrons in the
exact solution of Falicov-Kimball model with dynamical mean-field theory.  This problem was first investigated by
Brandt and Urbanek~\cite{brandt_urbanek} and by Janis~\cite{janis}.  The numerical
approach of Brandt and Urbanek was extended by Zlati\'c 
{\it et al.}~\cite{zlatic_review} and Freericks, Turkowski and 
Zlati\'c~\cite{freericks_turkowski_zlatic,freericks_turkowski_zlatic2,freericks_turkowski_zlatic3} within the original Brandt-Urbanek framework.
An alternative approach was taken by employing the numerical renormalization 
group by Anders and Czycholl~\cite{anders_czycholl}.  Some analytic work was 
completed by Liu~\cite{liu} which was related to early work on this 
problem~\cite{rutgers} and proceeded from the perspective of the X-ray edge 
problem~\cite{mahan,nozieres_dedominici}.
Here, we develop a new approach which is numerically more
tractable than other approaches and allows us to develop asymptotically exact
formulas for the Green's function in the time representation.

\begin{figure}[htb]
 \centerline{\includegraphics [width=2.0in, angle=0]  {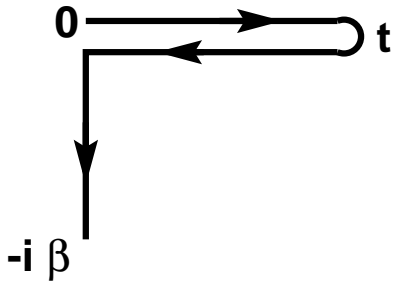}}
\caption[]{ The Kadanoff-Baym-Keldysh contour.  The contour starts at time 0, moves along the real time axis to time $t$, moves back along the real axis to time 0, then extends down the imaginary axis to time $-i\beta$.
}
\label{fig: contour}
\end{figure}

The single-site Hamiltonian of the Falicov-Kimball model~\cite{falicov_kimball}
involves two types of electrons---conduction electrons, denoted by the letter 
$d$ and localized electrons, denoted by the letter $f$
\begin{align}
 H_{\rm loc} &= E_f n_f + U n_d n_f -\mu\left(n_d + n_f \right)= H_0 (1-n_f) + H_1 n_f,\\
 H_0 &= -\mu n_d ,\\
 H_1 &= E_f - \mu + (U - \mu) n_d ,
\end{align}
where $n_d=d^\dag d$ and $n_f=f^\dag f$ are the number operators for the $d$- and $f$ electrons, respectively, $U$ is the on-site Coulomb interaction between the $d$ and $f$ electrons, and $E_f$ is the site-energy of the $f$ state. The full Hamiltonian on the lattice includes a repeat of this local Hamiltonian for each lattice site and a hopping of the conduction electrons between nearest-neighbor sites. The density matrix of the single-impurity problem is equal to
\begin{equation}
 \rho=e^{-\beta H_{\rm loc}} \mathcal{T}_c\exp\left\lbrace -i\int\nolimits_c dt'\int\nolimits_c dt''
               d^\dag(t')\lambda_c(t',t'') d(t'') \right\rbrace \frac{1}{\mathcal{Z}},
\end{equation}
where the time-ordering and integration are performed over the Kadanoff-Baym-Keldysh contour~\cite{kadanoff_baym,keldysh} and $\beta=1/T$ is the inverse temperature; the Kadanoff-Baym-Keldysh contour is shown in Fig.~\ref{fig: contour}---it starts at $t=0$, runs out along the real axis to a maximal time $t_{\rm max}$, then returns along the real axis back to $t=0$, and finally runs along the negative imaginary axis down to $-i\beta$.
Here, the dynamical mean-field $\lambda_c(t',t'')$ and chemical potential $\mu$ are taken from the equilibrium solution of the conduction electron problem via dynamical mean-field theory~\cite{freericks_review}. In other words, the dynamical mean field satisfies
\begin{equation}
 \lambda_c(t,t')=-\frac{i}{\pi}\int_{-\infty}^{\infty}d\omega{\rm Im}\lambda(\omega)e^{-i\omega(t-t')}[f(\omega)-\Theta_c(t,t')],
\end{equation}
where $\lambda(\omega)$ is the ordinary dynamical mean field on the real axis, extracted from the dynamical
mean-field theory solution of the model, $f(\omega)=1/[1+\exp(\beta\omega)]$ is the Fermi-Dirac distribution function,
and $\Theta_c(t,t')$ is the Heaviside function on the contour, equal to 1 when $t$ is ahead of $t'$ on the contour, equal to 0 when $t$ is behind $t'$ on the contour and equal to $1/2$ when $t=t'$ on the contour (note that this definition of the $\lambda_c$ field is $i$ times the definition
in Ref.~\cite{freericks_review}).
The partition function for the single-site problem contains two terms:
\begin{align}
 \mathcal{Z}=\Tr e^{-\beta H_{\rm loc}} \mathcal{T}_c\exp\left\lbrace -i\int\nolimits_c dt'\int\nolimits_c dt''
               d^\dag(t')\lambda_c(t',t'') d(t'') \right\rbrace =\mathcal{Z}_{0}+\mathcal{Z}_{1},
\end{align}
which correspond to the different occupations of the $f$-particles, as follows:
\begin{align}
 \mathcal{Z}_{0}&=\left[1+e^{\beta\mu} \right]\prod_m\left(1-\frac{\lambda_m}{i\omega_m+\mu} \right),
    & n_f=0; \\
 \mathcal{Z}_{1}&=e^{\beta(\mu-E_f)}\left[1+e^{\beta(\mu-U)} \right]\prod_m\left(1-\frac{\lambda_m}{i\omega_m+\mu-U} \right),
    & n_f=1.
\end{align}
Here, $i\omega_m=i\pi T(2m+1)$ is the fermionic Matsubara frequency and 
\begin{equation}
 \lambda_m=\int_0^\beta d\tau e^{i\omega_m\tau}\lambda(\tau)
\end{equation}
is the dynamical mean field evaluated at the $m$th Matsubara frequency [the $\lambda_c$ field is a function
only of the difference of the two time arguments when both times lie on the imaginary part of the Kadanoff-Baym-Keldysh contour and we use the notation $\lambda(\tau)=-i\lambda_c(-i\tau,0)$, which corresponds to the conventional definition of the dynamical mean field for imaginary time].

\section{Real-time Green's functions}

We consider the contour ordered Green's function for the $f$-electrons (with times $t$ and $t'$ both lying on the contour)
\begin{equation}
 G_f^c(t,t')=-i\left\langle\mathcal{T}_c f(t) f^\dag(t') \right\rangle ,
\end{equation}
where the time ordering is taken along the contour, the angle brackets denote taking a trace (weighted by
the density matrix for the equilibrium system), and the time dependence of the fermionic operators is with respect to the local Hamiltonian $H_{\rm loc}$.
Depending on the location of the time arguments on the contour, we obtain different real-time Green's functions:\\
the greater Green's function
\begin{equation}
 G_f^{>}(t-t')=-i\left\langle f(t) f^\dag(t') \right\rangle = -i \frac{1}{\mathcal{Z}}\sum_{pq}
   e^{-\beta\varepsilon_p}
         \left\langle p\left| f\right| q\right\rangle 
         \left\langle q\left| {f^{\dag}} \right| p\right\rangle 
         e^{i(\varepsilon_p-\varepsilon_q)(t-t')};
\end{equation}
the lesser Green's function
\begin{equation}
 G_f^{<}(t-t')=i\left\langle f^\dag(t') f(t) \right\rangle = i \frac{1}{\mathcal{Z}}\sum_{pq}
   e^{-\beta\varepsilon_q} 
         \left\langle q\left| {f^{\dag}} \right| p\right\rangle 
         \left\langle p\left| f\right| q\right\rangle 
         e^{i(\varepsilon_p-\varepsilon_q)(t-t')};
\end{equation}
the time-ordered Green's function
\begin{equation}
 G_f^t(t-t')=-i\left\langle\mathcal{T}_t f(t) f^\dag(t') \right\rangle 
    =\Theta(t-t')G_f^{>}(t-t') + \Theta(t'-t)G_f^{<}(t-t');
\end{equation}
and the antitime-ordered Green's function
\begin{equation}
 G_f^{\bar t}(t-t')=-i\left\langle\mathcal{T}_{\bar t} f(t) f^\dag(t') \right\rangle
    =\Theta(t'-t)G_f^{>}(t-t') + \Theta(t-t')G_f^{<}(t-t').
\end{equation}
Here $|p\rangle$ and $|q\rangle$ denote many-body eigenstates of the local Hamiltonian with eigenvalues
$\varepsilon_p$ and $\varepsilon_q$, respectively, and $\Theta(t)$ is the ordinary Heaviside function.
We are also interested in the retarded and advanced Green's functions,
\begin{align}
 G_f^r(t-t')&=-i\Theta(t-t')\left\langle \left[f(t), f^\dag(t')\right]_+ \right\rangle
 =\Theta(t-t')\left[G_f^{>}(t-t')-G_f^{<}(t-t') \right] ,\\
 G_f^a(t-t')&=i\Theta(t'-t)\left\langle \left[f(t), f^\dag(t')\right]_+ \right\rangle
 =\Theta(t'-t)\left[G_f^{<}(t-t')-G_f^{>}(t-t') \right] ,
\end{align}
which are easily constructed from taking combinations of the lesser and greater Green's functions (note the symbol $[...]_+$ denotes the anticommutator of the two operators in the brackets).
In equilibrium, the Green's functions depend only on the time difference of the real-time variables and the greater and lesser Green's functions also satisfy
\begin{align}\label{minus_t}
 \left[G_f^{>}(t-t')\right]^*=-G_f^{>}(t'-t), \qquad \left[ G_f^{<}(t-t')\right]^* =-G_f^{<}(t'-t).
\end{align}

It is often useful to represent quantities in terms of frequencies, instead of time. The Fourier transforms of the greater and lesser Green's functions (to frequency space) are equal to:
\begin{equation}
 G_f^{>}(\omega)=-2\pi i [1-f(\omega)]A_f(\omega),\quad
 G_f^{<}(\omega)=2\pi i f(\omega)A_f(\omega),
\end{equation}
where
\begin{equation}
 A_f(\omega)=\frac{1}{\mathcal{Z}}\sum_{pq}
   [e^{-\beta\varepsilon_p}+e^{-\beta\varepsilon_q}]
         \left\langle p\left| f\right| q\right\rangle 
         \left\langle q\left| {f^{\dag}} \right| p\right\rangle 
         \delta(\omega+\varepsilon_p-\varepsilon_q)
\end{equation}
is the $f$-electron spectral density. The corresponding Fourier transforms of the retarded and time-ordered Green's functions are the following:
\begin{equation}
 G_f^r(\omega)= \int_{-\infty}^{+\infty}d\omega'
           \frac{A_f(\omega')}{\omega-\omega'+i\delta},\quad
 \Img G_f^r(\omega) =-\pi A_f(\omega),
\label{ImGvsIw}
\end{equation}
and
\begin{equation}
 G_f^t(\omega)=\Real G_f^r(\omega) + i \Img G_f^r(\omega) \tanh \frac{\beta\omega}{2}.
\end{equation}

We use the Kadanoff-Baym-Keldysh approach to calculate the real-time Green's function because there is
no simple way to perform the analytical continuation of the Matsubara frequency Green's function to the
real axis~\cite{brandt_urbanek}. We are interested in the retarded Green's function, so
we consider the greater Green's function for positive time $t>0$
\begin{equation}\label{G_>}
 G_f^{>}(t)=-i\frac{1}{\mathcal{Z}}\Tr \left \{ e^{-\beta H_{\rm loc}} \mathcal{T}_c\exp\left [ -i
   \int\nolimits_c dt'\int\nolimits_c dt'' d^\dag(t')\lambda_c(t',t'') d(t'') \right ]
    f(t) f^\dag(0)\right \}
\end{equation}
and the lesser Green's function for negative time $t<0$ (or $\bar t=-t>0$) 
\begin{equation}\label{G_<}
 G_f^{<}(t)=i\frac{1}{\mathcal{Z}}\Tr \left \{ e^{-\beta H_{\rm loc}} \mathcal{T}_c\exp\left [ -i
   \int\nolimits_c dt'\int\nolimits_c dt'' d^\dag(t')\lambda_c(t',t'') d(t'') \right ]
   f^\dag(\bar t) f(0)\right \};
\end{equation}
we use the fact that $f^\dagger(0)f(t)$ can be replaced by $f^\dagger(\bar t)f(0)$ in the trace due to the time-translation invariance that we have in equilibrium.
The greater and lesser Green's functions can be found for other time values by using the relations in equation~(\ref{minus_t}).

The first step in solving for the Green's function is to write equations of motion for the $f$-operators:
\begin{align}
 \frac{df(t)}{dt}&=-i\left[ U n_d(t) + E_f - \mu \right] f(t),\\
 \frac{df^\dag(\bar t)}{d\bar t}&=i\left[ U n_d(\bar t) + E_f - \mu \right] f^\dag(\bar t),
\end{align}
and substitute their solutions into equations~(\ref{G_>}) and (\ref{G_<}), which yields the following expressions for the greater Green's function ($t>0$)
\begin{align}%\label{G_>}
 G_f^{>}(t)=&-i\frac{1}{\mathcal{Z}}\Tr \left \{ e^{-\beta H_{\rm loc}} \mathcal{T}_c\exp\left [ -i
   \int\nolimits_c dt'\int\nolimits_c dt'' d^\dag(t')\lambda_c(t',t'') d(t'') \right.\right .\nonumber\\
   &\left.\left. {}-i \int\nolimits_c dt' U_c(t,t') n_d(t') -i( E_f - \mu)t \right ]
    f(0) f^\dag(0)\right \}
\nonumber
\end{align}
and for the lesser Green's function ($\bar t=-t>0$)
\begin{align}%\label{G_>}
 G_f^{<}(t)=&i\frac{1}{\mathcal{Z}}\Tr \left \{ e^{-\beta H_{\rm loc}} \mathcal{T}_c\exp\left [ -i
   \int\nolimits_c dt'\int\nolimits_c dt'' d^\dag(t')\lambda(t',t'') d(t'') \right.\right.\nonumber\\
   &\left.\left. {}+i \int\nolimits_c dt' U_c(\bar t,t') n_d(t') +i( E_f - \mu)\bar t \right ]
   f^\dag(0) f(0)\right \},
\nonumber
\end{align}
where $U_c(t,t')$ is a new time-dependent field, which is nonzero only on the upper branch of the Kadanoff-Baym-Keldysh contour and is equal to $U$ for $t'\in[0,t]$ and is zero otherwise. It is because this $U_c$ field is not time-translation invariant that we need to use the Kadanoff-Baym-Keldysh formalism for the analytic continuation.
Now we take into account the fact that the $f$-particle occupation number $n_f=f^\dag f$ is a conserved quantity of the Hamiltonian; for the greater Green's function we have a projection onto states without $f$-particles ($n_f=0$) while for the lesser Green's function we have a projection onto states with one $f$-particle ($n_f=1$), which yields:
\begin{align}\label{G_>_sub}
 G_f^{>}(t)&=-i\frac{1}{\mathcal{Z}}e^{-i(E_f - \mu)t}\nonumber\\
&\times\Tr \left \{ e^{\beta\mu n_d}
   \mathcal{T}_c\exp\left [ -i\int\nolimits_c dt'\int\nolimits_c dt''
         d^\dag(t')\lambda_c(t',t'') d(t'') 
   -i \int\nolimits_c dt' U_c(t,t') n_d(t') \right ]\right \}
\end{align}
and
\begin{align}\label{G_<_sub}
 G_f^{<}(t)=&i\frac{1}{\mathcal{Z}} e^{\beta(\mu-E_f)}e^{i(E_f - \mu)\bar t}
\\ \nonumber
   &\times\Tr \left\{ e^{\beta(\mu-U)n_d} \mathcal{T}_c\exp\left [ -i
   \int\nolimits_c dt'\int\nolimits_c dt'' d^\dag(t')\lambda_c(t',t'') d(t''),
   +i \int\nolimits_c dt' U_c(\bar t,t') n_d(t') \right] \right \},
\end{align}
where the trace is now over the $d$-electrons only.

We next expand the contour-ordered-exponent into a series that we can ultimately sum exactly. To begin, we need to recall Wick's theorem, where the expectation value of any number of pairs of fermionic operators can be expanded in terms of products of the two-operator contractions
\begin{align}\label{rg0}
  \stackrel{\unitlength=1em\line(0,-1){.3}\vector(-1,0){1.4}\line(-1,0){.8}\line(0,-1){.3}\qquad}
  {d(t_{1}) d^{\dag}(t_2)}& = i g_{\alpha}(t_{1},t_2),
\\ \nonumber
g_{\alpha}(t_{1},t_2)&=-i\left\langle \mathcal{T}_{c} d(t_{1}) d^{\dag}(t_2) \right\rangle_{\alpha}
= i e^{-i\epsilon_{\alpha}(t_{1}-t_2)}\left[ f(\epsilon_{\alpha}) - \Theta_c(t_{1},t_2) \right]
\end{align}
when the Hamiltonian is quadratic in the fermionic operators.
Here we have $\epsilon_{0}=-\mu$ for sites with $n_f=0$ ($\alpha=0$) and $\epsilon_{1}=U-\mu$ for sites with $n_f=1$ ($\alpha=1$).  The traces in equations~(\ref{G_>_sub}) and (\ref{G_<_sub}) can then be written in the following diagrammatic expression
\begin{equation}\label{action_diag}
\left(1+e^{-\beta\epsilon_{\alpha}}\right)
 \exp \left\{ \raisebox{-12pt}[15pt][12pt]{\includegraphics[scale=0.7]{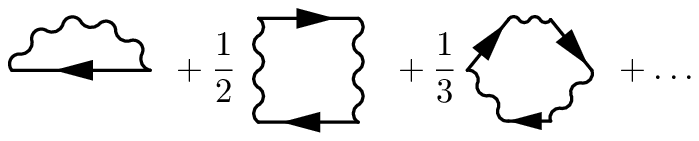}}
\right\},
\end{equation}
where the arrows denote the zero-order Green functions $g_{\alpha}(t',t'')$ in equation~(\ref{rg0}) and the wavy lines denote a generalized time-dependent hopping ($\lambda$-field) 
\begin{equation}
\tilde\lambda_t(t',t'') = \lambda(t',t'') + U_c(t,t')\delta_c(t',t'') 
%\tilde\lambda_{\bar t}(t',t'') = \lambda(t',t'') - U_c(\bar t,t')\delta_c(t',t'') 
\end{equation} 
for the greater Green's function and we have to replace 
\begin{equation}\label{U_subst}
 U_c(t,t')\to-U_c(\bar t,t')
\end{equation}
for the lesser Green's function.
% As a rule, in order to sum up an infinite series in (\ref{action_diag}) one have to introduce an interaction strength $U\to\alpha U$ ($\alpha\in[0,1]$) and use standart tricks. As a result, final expressions will contain integral over interaction strength $\alpha$. 

%Here we use similar approach which allows to remove additional quantities from the final expressions. 
Motivated by approaches that involve the integration over a coupling constant, we modify our time-dependent field by introducing a dependence on some new parameter that we denote by $x$
\begin{equation}
 U_c(t,t')\to U_c(t,t'|x),
\end{equation}
with the following limiting behavior
\begin{equation}
 U_c(t,t'|0)=0,\qquad
 U_c(t,t'|1)=U_c(t,t').
\end{equation}
 Next, we take the derivative of the diagrammatic series in equation~(\ref{action_diag}) with respect to $x$  and find:
\begin{equation}
\frac{d}{dx}\Bigl\{\ldots\Bigr\} = -\int_c dt' \frac{dU_c(t,t'|x)}{dx} G_{U}^{(\alpha)}(t',t'|x),
\end{equation}
where we introduced a parameter-dependent Green's function $G_{U}^{(\alpha)}(t',t''|x)$, which is the solution of the following Dyson equation:
\begin{align}\label{rDyson_tau}
G_{U}^{(\alpha)}(t',t''|x) =& g_{\alpha}(t',t'') + 
\int\nolimits_c dt_1 \int\nolimits_c dt_2\, g_{\alpha}(t',t_1) \tilde\lambda_t(t_1,t_2|x)
G_{U}^{(\alpha)}(t_2,t''|x)
\\
 =& g_{\alpha}(t',t'') + 
\int\nolimits_c dt_1 \int\nolimits_c dt_2\, g_{\alpha}(t',t_1) \lambda_c(t_1,t_2)
G_{U}^{(\alpha)}(t_2,t''|x)
\nonumber
\\ 
&+ \int\nolimits_c dt_1\, g_{\alpha}(t',t_1) U_c(t,t_1|x)
G_{U}^{(\alpha)}(t_1,t''|x).
\nonumber
\end{align}
This expression holds for the greater Green's function ($\alpha=0$). For the lesser Green's function, one has to use the substitution in equation~(\ref{U_subst})  and set $\alpha=1$. Finally, we find explicit expressions for the greater
\begin{equation}\label{rFvsG}
 G_f^{>}(t)=-i w_0 \exp\left\lbrace -i(E_f - \mu)t
     - \int_0^1dx\int\nolimits_c dt_1\, \frac{dU_c(t,t_1|x)}{dx} G_{U}^{(0)}(t_1,t_1|x)\right\rbrace
\end{equation}
and lesser
\begin{equation}\label{rFvsG_l}
 G_f^{<}(t)=i w_1 
  \exp\left\lbrace i(E_f - \mu)\bar t + \int_0^1dx \int\nolimits_c dt_1\, \frac{dU_c(\bar t,t_1|x)}{dx}
         G_{U}^{(1)}(t_1,t_1|x)\right\rbrace ,
\end{equation}
Green functions, where
\begin{equation}
 w_0=\langle 1- n_f\rangle=\frac{\mathcal Z_{0}}{\mathcal Z_{0}+\mathcal Z_{1}}, \qquad 
w_1=\langle n_f\rangle=\frac{\mathcal Z_{1}}{\mathcal Z_{0}+\mathcal Z_{1}}
\end{equation}
are the average densities of sites without ($w_0$) and with ($w_1$) $f$-electrons.

We can rewrite equation~(\ref{rDyson_tau}) in the following two forms:
\begin{equation}\label{rDyson_aux}
G_{U}^{(\alpha)}(t',t''|x) = g_{U{\alpha}}^{aux}(t',t''|x) + 
\int\nolimits_c dt_1 \int\nolimits_c dt_2\, g_{U{\alpha}}^{aux}(t',t_1|x) \lambda_c(t_1,t_2) G_{U}^{(\alpha)}(t_2,t''|x),
\end{equation}
or
\begin{equation}\label{rDyson_0}
G_{U}^{(\alpha)}(t',t''|x) = G_{\alpha}(t',t'') + \int\nolimits_c dt_1\, G_{\alpha}(t',t_1) U_c(t,t_1|x) G_{U}^{(\alpha)}(t_1,t''|x).
\end{equation}
The first form emphasizes the lambda field as an effective self-energy, while the second emphasizes the $U$ field as the effective self-energy. 
In the top equation, we introduced the auxiliary Green's function
\begin{equation}
g_{U{\alpha}}^{aux}(t',t''|x) = g_{\alpha}(t',t'') + \int\nolimits_c dt_1\, g_{\alpha}(t',t_1) U_c(t,t_1|x) g_{U{\alpha}}^{aux}(t_1,t''|x)
\end{equation}
as first defined by Brandt and Urbanek~\cite{brandt_urbanek}, while
we introduced the bare time-ordered Green's function for the effective medium
\begin{equation}\label{rGF_0}
G_{\alpha}(t',t'') = g_{\alpha}(t',t'') + 
\int\nolimits_c dt_1 \int\nolimits_c dt_2\, g_{\alpha}(t',t_1) \lambda_c(t_1,t_2) G_{\alpha}(t_2,t''),
\end{equation}
in the bottom equation. The bare time-ordered Green's function for the effective medium
can be determined directly on the real axis via
\begin{equation}\label{rGF_0_ex}
G_{\alpha}(t',t'') = -\frac{i}{\pi}\int_{-\infty}^{+\infty} d\omega\, \Img \mathcal G_{\alpha}(\omega) e^{-i\omega (t'-t'')} \left[ f(\omega) - \Theta_c(t',t'') \right] ,
\end{equation}
with
\begin{equation}
\mathcal G_{\alpha}(\omega)=\frac{1}{\omega+i\delta-\epsilon_{\alpha}-\lambda(\omega+i\delta)}
\end{equation}
being the retarded effective medium on the real frequency axis.

The first representation (\ref{rDyson_aux}) was originally used by Brandt and Urbanek~\cite{brandt_urbanek} and improved by Zlati\'c \textit{et al.}~\cite{zlatic_review} and Freericks, Turkowski and Zlati\'c~\cite{freericks_turkowski_zlatic} (see for review Freericks and Zlati\'c~\cite{freericks_review}) and requires the calculation of continuous matrix operator determinants over the entire contour.
In what follows, we shall instead use the second representation in equation~(\ref{rDyson_0}) because in equations~(\ref{rFvsG}) and (\ref{rDyson_0}) all times in $G_{U}^{\alpha}(t_1,t_1|x)$ are only on the real axis ($0\le t_1\le t$) [or ($0\le t_1\le\bar t$)] and we can replace the contour integral in equation~(\ref{rDyson_0}) by an integral over the upper real time branch of the Kadanoff-Baym-Keldysh contour where the time-dependent field $U_c(t,t_1|x)$ is nonzero
\begin{equation}\label{rDyson_0r}
G_{U}^{\alpha}(t',t''|x) = G_{\alpha}(t'-t'') + \int_{-\infty}^{+\infty} dt_1\, G_{\alpha}(t'-t_1)
U(t,t_1|x) G_{U}^{\alpha}(t_1,t''|x).
\end{equation}
Integrals over the entire Kadanoff-Baym-Keldysh contour survive only in the definition of $G_{\alpha}(t)$ in equation~(\ref{rGF_0}), which can instead be calculated directly on the real axis by picking $t$ and $t'$ on the upper branch of the contour in equation~(\ref{rGF_0_ex})
\begin{equation}\label{rGF_0_exr}
G_{\alpha}(t'-t'') = -\frac{i}{\pi}\int_{-\infty}^{+\infty} d\omega\, \Img \mathcal G_{\alpha}(\omega) e^{-i\omega (t'-t'')} \left[ f(\omega) - \Theta(t'-t'') \right] .
\end{equation}
Also, for further discussions we will need its Fourier transform
\begin{equation}\label{rGF_0_FT}
G_{\alpha}(\omega)
=\int_{-\infty}^{+\infty}dt\, e^{i\omega t } G_{\alpha}(t)
= \Real \mathcal G_{\alpha}(\omega) + i\tanh\frac{\beta\omega}{2} \Img \mathcal G_{\alpha}(\omega).
\end{equation}

We are now ready to summarize the final formulas for the Green's function by evaluating all of the terms in equation~(\ref{rDyson_0}) and then performing the integration over the parameter $x$.  We begin with the formal matrix solution for $G_U^\alpha(t',t''|x)$
\begin{equation}
G_{U}^{\alpha}(t',t''|x)=\int\nolimits_c d\tilde t \,
\left[\left\| \delta_c(t_1,t_2) - G_{\alpha}(t_1,t_2)U_c( t,t_2|x) \right\|^{-1}\right]_{t'\tilde t} 
G_{\alpha}(\tilde t,t'').
\end{equation}
Here $\delta_c(t,t')$ is the Dirac delta function along the contour defined by $\int_c dt'\delta_c(t,t')f(t')=f(t)$.
After substituting this result into equation~(\ref{rFvsG}) and then integrating over the parameter $x$, we obtain for the greater Green's function the following results:
\begin{align}
 G_f^{>}(t)=&-iw_0\exp\bigg\lbrace -i(E_f - \mu)t
\\
     &- \int_0^1 dx\int\nolimits_c dt'\, \int\nolimits_c d\tilde t \,
\left[\left\| \delta_c(t_1,t_2) - G_{0}(t_1,t_2)U_c( t,t_2|x) \right\|^{-1}\right]_{t'\tilde t} 
G_{0}(\tilde t,t') \frac{dU_c(t,t'|x)}{dx} \bigg\rbrace
\nonumber
\\
=&-iw_0\exp\left\lbrace -i(E_f - \mu)t
     - \int_0^1 dx \Tr_c \left [
\left\| \mathbf{I} - \mathbf{G}_{0}\mathbf{U}_c(t|x) \right\|^{-1}
\mathbf{G}_{0} \frac{d\mathbf{U}_c(t|x)}{dx} \right ]\right\rbrace
\nonumber
\\
=&-iw_0\exp\left\lbrace -i(E_f - \mu)t
     + \ln \det\nolimits_c
\left\| \mathbf{I} - \mathbf{G}_{0}\mathbf{U}_c(t) \right\|
\right\rbrace
\nonumber
\\
=&-iw_0 e^{-i(E_f - \mu)t}
      \det\nolimits_c
\left\| \delta_c(t_1,t_2) - G_{0}(t_1,t_2) U_c(t,t_2) \right\|,
\nonumber
\end{align}
where $\Tr_c$ denotes the trace of a continuous matrix operator expressed as a line integral over the contour and $\det\nolimits_c$ is the determinant of the continuous matrix operator defined on the contour. However, in our case, the time-dependent field $U_c(t,t_2)$ (and hence all nondiagonal components) are nonzero only on the upper real time branch of the contour for $0\le t_2\le t$, which allows us to reduce the problem to the determinant of a continuous matrix operator over the finite real time interval:
\begin{equation}
 G_f^{>}(t)=-iw_0 e^{-i(E_f - \mu)t}
      \det\nolimits_{[0,t]}
\left\| \delta(t_1-t_2) - G_{0}(t_1-t_2) U \right\|.
\end{equation}
{\it It is the restriction to this finite real-time interval that makes the numerical calculations much more efficient than other methods used previously.}
Note that all of the temperature dependence is in the Green's function $G_0$, which is the time-ordered Green's function for the effective medium of the dynamical mean-field theory.  

In a similar fashion, we can derive the expression for the lesser Green's function ($\bar t=-t>0$):
\begin{equation}
 G_f^{<}(t)=iw_1 e^{i(E_f - \mu)\bar t}
      \det\nolimits_{[0,\bar t]}
\left\| \delta(t_1-t_2) + G_{1}(t_1-t_2) U \right\|.
\end{equation}
Here, the Green's function $G_{1}(t_1-t_2)$ is defined in equation~(\ref{rGF_0_exr}) and is the time-ordered Green's function for the effective medium when there is an $f$-electron on the impurity site. 

When we perform numerical calculations to evaluate the continuous matrix determinants, we first discretize the time contour with a discretization step $\Delta t$, and then discretize the continuous matrix operator into a discrete matrix, which can be manipulated by standard computational libraries like LAPACK. The discretized representation for the Green's functions is then
\begin{equation}
 G_f^{>}(t)=-iw_0 e^{-i(E_f - \mu)t}
       \det\nolimits_{[0,t]}\left\|\delta_{ij}-W_i G_{0}(t_i-t_j) U\right\| 
\label{eq: g_greater_discrete}
\end{equation}
and
\begin{equation}
 G_f^{<}(t)=iw_1 
    e^{i(E_f - \mu)\bar t} 
    \det\nolimits_{[0,\bar t]}\left\|\delta_{ij}+W_i G_{1}(t_i-t_j) U\right\|,
\label{eq: g_lesser_discrete}
\end{equation}
where $W_i$ are the quadrature weights for the discrete set of times $\{t_i\}$; for a uniform grid $W_i=W=\Delta t$, we have to calculate the determinant of a Toeplitz matrix. This will allow us to use Szeg\"o's theorem and the Wiener-Hopf approach to find analytic expressions of the continuous matrix operator determinant which are accurate for long times.  Note, that while it appears that these expressions are equally valid for all temperatures, the numerical solution, in the $\Delta t\rightarrow 0$ limit, becomes more difficult at lower temperatures.

The retarded Green's function is found from the greater and lesser Green functions via
\begin{align}
 G_f^r(t)&=\Theta(t)\left\{G_f^{>}(t)-G_f^{<}(t) \right\}.
\end{align}
Using the relation in equation~(\ref{minus_t}), then yields
\begin{align}
 G_f^r(t)&=-i\Theta(t) e^{-i(E_f - \mu)t} \left\{
   w_0 \det\nolimits_{[0,t]} \left\| \mathbf{I} - \mathbf{G}_{0} U \right\|
+ w_1 \left(\det\nolimits_{[0,t]} \left\| \mathbf{I} + \mathbf{G}_{1} U \right\| \right)^*
\right\}
\end{align}
for real times, and
\begin{align}\label{Gr_w_total}
 G_f^r(\omega)&=-i\int_0^{+\infty}dt e^{i(\omega + \mu -  E_f)t}
 \left\{
w_0 \det\nolimits_{[0,t]} \left\| \mathbf{I} - \mathbf{G}_{0} U \right\|
+ w_1 \left(\det\nolimits_{[0,t]} \left\| \mathbf{I} + \mathbf{G}_{1} U \right\| \right)^*
\right\}
\end{align}
for real frequencies.

\section{The Wiener-Hopf sum equation approach and Szeg\"o's theorem}

We follow closely the development of the Wiener-Hopf sum equation approach in McCoy and Wu~\cite{mccoy_wu}, which is based on the original integral equation development by Wiener and Hopf~\cite{wiener_hopf}, as summarized in Krein~\cite{krein}.  The Wiener-Hopf integral equation approach has been used widely in the X-ray edge problem and in the Falicov-Kimball model~\cite{janis} primarily at $T=0$; we will show that the Wiener-Hopf sum equation approach is very powerful for finite temperature calculations.

Before we can apply this technique to the problem of calculating the $f$-electron spectral function, we must begin with some mathematical preliminaries (further details can be found in Ref.~\cite{mccoy_wu}).  Consider a sequence of numbers $c_n$ which satisfy $\sum_{n=-\infty}^\infty|c_n|<\infty$ and the set of $N+1$ linear equations
\begin{equation}
 \sum_{m=0}^Nc_{n-m}x^{(N)}_m=y_n,
\label{eq: wh1}
\end{equation}
for $0\le n\le N$ with $c_n$ and $y_n$ known.  We will eventually be taking the limit where $N\rightarrow\infty$, so we will require $\sum_{n=-\infty}^\infty|y_n|<\infty$ too. Note that the matrix $c_{n-m}$ that appears in equation~(\ref{eq: wh1}) is a Toeplitz matrix, because it depends only on the difference $n-m$ and not on $n$ and $m$ separately. Our job is to find the solution $x^{(N)}_m$ to these equations. We define 
\begin{equation}
 x_n^{(N)}=y_n=0
\end{equation}
for $n<0$ and $n>N$. In addition, we need to define two more sets of coefficients:
\begin{displaymath}
 u_n=\left \{ 
% use packages: array
\begin{array}{ll}
\sum_{m=0}^Nc_{N+n-m}x^{(N)}_m & {\rm for}~n>0 \\ 
0 & {\rm for}~n\le 0
              \end{array}
\right .
\label{eq: u_def}
\end{displaymath}
and
\begin{displaymath}
 v_n=\left \{ 
% use packages: array
\begin{array}{ll}
\sum_{m=0}^Nc_{-n-m}x^{(N)}_m & {\rm for}~n>0 \\ 
0 & {\rm for}~n\le 0
              \end{array}
\right ..
\label{eq: v_def}
              \end{displaymath}
Next, we relax the condition on equation~(\ref{eq: wh1}) to apply it for all $n$ and we allow the sum over $m$ to extend from $-\infty$ to $\infty$ (although all cases have just a finite number of terms because $x_m^{(N)}$ is potentially nonzero for only $N+1$ terms). The result is
\begin{equation}
 \sum_{m=-\infty}^{+\infty} c_{n-m}x^{(N)}_m=y_n+\Theta(n> N)u_{n-N}+\Theta(n<0)v_{-n}.
\label{eq: wh2}
\end{equation}
We consider a variable $\xi=\exp[i\theta]$ which lies on the unit circle, so that $|\xi|=1$.  We multiply both sides of equation~(\ref{eq: wh2}) by $\xi^n$ and sum over $n$ to find
\begin{equation}
 C(\xi)X(\xi)=Y(\xi)+U(\xi)\xi^N+V(\xi^{-1}),
\label{eq: wh3}
\end{equation}
with
\begin{align}\label{eq: wh_fcn_defs}
&C(\xi)=\sum_{n=-\infty}^{+\infty}c_n\xi^n,\quad
X(\xi)=\sum_{n=0}^Nx_n^{(N)}\xi^n,\quad 
Y(\xi)=\sum_{n=0}^N y_n\xi^n,
\\
\nonumber
&U(\xi)=\sum_{n=1}^\infty u_n\xi^n,\quad\textrm{and}\quad
V(\xi)=\sum_{n=1}^\infty v_n\xi^n.
\end{align}
It may appear that we have made the problem more complicated, because we have replaced a problem with one unknown $X(\xi)$ by a problem with three unknowns $X(\xi)$, $U(\xi)$ and $V(\xi)$.  But it turns out that this more complicated problem can actually be solved by carefully analyzing its analytic structure.

The crucial step of Wiener and Hopf is to find a unique factorization of $C(\xi)$ into two factors, one analytic
inside the unit circle and continuous on the circle, and the other analytic outside the unit circle and continuous on
the circle.  Before we show how to do this, we need some more mathematical definitions.  We call a function a $+$ function if it can be expanded as a Laurent series for $|\xi|=1$ ($\sum_{n=0}^\infty a_n\xi^n$) and the coefficients satisfy $\sum_{n=0}^\infty |a_n|<\infty$.  In this case, the function is analytic within the unit circle and continuous on the unit circle.  Similarly, we call a function a $-$ function if it can be expanded in a Laurent series for $|\xi|=1$
($\sum_{n=-\infty}^{-1}a_n\xi^n$) and the coefficients also satisfy $\sum_{n=-\infty}^{-1} |a_n|<\infty$.  This function is analytic outside the unit circle and continuous on the unit circle; it vanishes as $|\xi|\rightarrow\infty$.  Note that any function defined on the unit circle via a Laurent expansion $f(\xi)=\sum_{n=-\infty}^\infty a_n\xi^n$ with $\sum_{n=-\infty}^\infty |a_n|<\infty$ can obviously be uniquely decomposed into the sum of a $+$ and a $-$ functions via $f(\xi)=f_+(\xi)+f_-(\xi)$.

The function $C(\xi)$ is factorized in a so-called canonical factorization, where
\begin{equation}
 C(\xi)=P(\xi)^{-1}Q(\xi^{-1})^{-1},
\label{eq: c_fact}
\end{equation}
and both $P(\xi)$ and $Q(\xi)$ are nonzero for $|\xi|\le 1$; the factorization is made unique by requiring $Q(0)=1$. The functions $P$ and $Q$ can then be found in terms of $+$ and $-$ functions as
\begin{equation}
 P(\xi)=e^{G_+(\xi)}\quad{\rm and}\quad Q(\xi^{-1})=e^{G_-(\xi)},
\label{eq: p_q}
\end{equation}
which automatically satisfies $G_-(\infty)=0$ [and $Q(0)=1$] and makes both $P(\xi)$ and $Q(\xi)$ nonvanishing inside the unit circle. An explicit calculation of $G_{\pm}(\xi)$ is nontrivial, but it is easy to write down a formal solution via
\begin{equation}
 G_+(\xi)=-\left [ \ln C(\xi)\right ]_+\quad{\rm and}\quad G_-(\xi)=-\left [ \ln C(\xi)\right ]_-,
\label{eq: gpm_def}
\end{equation}
where the $\pm$ subscripts denote a restriction of the Laurent series for the logarithm to its nonnegative or negative power terms, respectively.  Note that the condition for properly defining the $+$ and $-$ functions requires the logarithm to be single-valued after $\xi$ winds around the unit circle, or, in other words, we require $\Ind C(\xi)=0$, where
\begin{equation}
 \Ind C(\xi)=\frac{1}{2i\pi}\left [ \ln C(e^{2i\pi})-\ln C(e^{i0})\right ].
\label{eq: index_def}
\end{equation}
We will also need to consider situations where $\Ind C(\xi)=-1$.  In this case, a new function $\bar C(\xi)=-\xi C(\xi)$ will have zero index, and the canonical factorization can be applied to $\bar C(\xi)$ instead.

We are now ready to solve equation~(\ref{eq: wh3}) by substituting in the factorization from equation~(\ref{eq: c_fact}).  Next, multiply both sides of the equation by $Q(\xi^{-1})$ to get
\begin{equation}
 [P(\xi)]^{-1}X(\xi)=Q(\xi^{-1})Y(\xi)+Q(\xi^{-1})U(\xi)\xi^N+Q(\xi^{-1})V(\xi^{-1}).
\label{eq: wh4}
\end{equation}
The term on the left hand side is a $+$ function, because $P(0)\ne 0$, and $[P(\xi)]^{-1}$ can be expanded in
a power series with just positive powers for $|\xi|<1$ and $X(\xi)$ is a $+$ function.  Similarly, $Q(\xi^{-1})V(\xi^{-1})$ is a $-$ function.   The other two terms are neither $+$ nor $-$ functions, but they can be uniquely decomposed into $+$ and $-$ function pieces.  We do this, and move all $+$ functions to the left hand side and all $-$ function pieces to the right hand side.  We are left with
\begin{eqnarray}
&~&[P(\xi)]^{-1}X(\xi)-\left [ Q(\xi^{-1})Y(\xi)\right ]_+-\left [Q(\xi^{-1})U(\xi)\xi^N\right ]_+
=\nonumber\\
&~&~~\quad\left [ Q(\xi^{-1})Y(\xi)\right ]_-+\left [Q(\xi^{-1})U(\xi)\xi^N\right ]_-+Q(\xi^{-1})V(\xi^{-1}).
\label{eq: wh5}
\end{eqnarray}
The left hand side of equation~(\ref{eq: wh5}) is an analytic function for $|\xi|<1$ and is continuous on the unit circle, the right hand side is an analytic function for $|\xi|>1$ and is continuous on the unit circle, and both functions agree on the unit circle, due to the equality, so we can define a function that is analytic in the entire complex plane, and hence is an entire function.  But this function vanishes as $|\xi|\rightarrow\infty$, and the only entire function that vanishes for large argument is the function that is identically equal to 0.  Hence we learn that
\begin{equation}
 X(\xi)=P(\xi)\left [ Q(\xi^{-1})Y(\xi)\right ]_++P(\xi)\left [Q(\xi^{-1})U(\xi)\xi^N\right ]_+
\label{eq: whfinal1}
\end{equation}
and
\begin{equation}
 V(\xi^{-1})=-\left [ Q(\xi^{-1})\right ]^{-1}\left \{ \left [ Q(\xi^{-1})Y(\xi)\right ]_-+\left [Q(\xi^{-1})U(\xi)\xi^N\right ]_-\right \}.
\label{eq: whfinal2}
\end{equation}
Using the fact that $X(\xi^{-1})\xi^N$ is a $+$ function [recall $X(\xi)$ is an order $N$ polynomial in $\xi$], we can substitute $\xi\rightarrow\xi^{-1}$ in equation~(\ref{eq: wh5}), multiply both sides of the equation by $\xi^N$, and then separate into the $+$ and $-$ function pieces to create another vanishing entire function, and learn that both
\begin{equation}
 X(\xi^{-1})\xi^N=Q(\xi)\left \{ \left [P(\xi^{-1})Y(\xi^{-1})\xi^N\right ]_++\left [ P(\xi^{-1})V(\xi)\xi^N\right ]_+\right \}
\label{eq: whfinal3}
\end{equation}
and
\begin{equation}
 U(\xi^{-1})=-\left [P(\xi^{-1})\right ]^{-1}\left \{ \left [ P(\xi^{-1})Y(\xi^{-1})\xi^N\right ]_-+\left [ P(\xi^{-1})V(\xi)\xi^N\right ]_-\right \}
\label{eq: whfinal4}
\end{equation}
hold.  These are all of the necessary relations we will need for determining the Toeplitz determinants.

Now we must take another mathematical interlude to state Szeg\"o's theorem.  This theorem determines how a Toeplitz determinant exponentially decays for large matrix size~\cite{szego} and even finds the constant prefactor for the exponential decay~\cite{szego2,szego3}.  The proof can be carried out using the Wiener-Hopf sum equation approach~\cite{mccoy_wu}, but it would take too much space for us to repeat it here, and it is rife with many technical mathematical manipulations that would take us away from the physical phenomena we wish to discuss here.  So we will instead simply state the result, and then show how to evaluate asymptotic expressions for the $f$-electron Green's function in real time.

If we consider the determinant $D_N$ of the $N\times N$ Toeplitz matrix defined by
\begin{displaymath}
 D_N=\left | 
\begin{array}{ccccc}
 c_0 & c_{-1} & c_{-2} & \hdots & c_{-N+1}\\
c_1 & c_0 & c_{-1}& \hdots & c_{-N+2}\\
c_2 & c_1 & c_0 & \hdots & c_{-N+3}\\
\vdots & \vdots & \vdots & \vdots & \vdots\\
c_{N-2} & c_{N-3} & c_{N-4} & \hdots & c_{-1} \\
c_{N-1} & c_{N-2} & c_{N-3}& \hdots & c_0
\end{array}
\right |,
\end{displaymath}
then Szeg\"o's theorem says that
\begin{equation}
 \lim_{N\rightarrow\infty} D_N=\exp\left [ N g_0 + \sum_{n=1}^\infty n g_n g_{-n}\right ],
\label{eq: szego}
\end{equation}
with
\begin{equation}
 g_n=\frac{1}{2\pi}\int_{-\pi}^{\pi}d\theta e^{-in\theta}\ln C(e^{i\theta}).
\label{eq: g_t_def}
\end{equation}
The conditions that $\Ind C(\xi)=0$, $C(\xi)$ is continuous on the unit circle, and $\sum_{n=-\infty}^{\infty}|c_n|<\infty$ are all required for the theorem to hold.

Now, with all of the mathematical preliminaries complete, we are ready to apply these techniques to finding the Toeplitz determinants needed for the $f$-electron Green's function. We need to evaluate the two determinants in equations~(\ref{eq: g_greater_discrete}) and (\ref{eq: g_lesser_discrete}).  We will show explicitly how to calculate the determinant for the greater Green's function.  Modifications for the lesser Green's function are straightforward.

\paragraph{Case 1: no winding.}

We start by discretizing the time axis $[0,t]$ with $N+1$ points given by $t_n=n\Delta t$, with $\Delta t=t/N$. Then we define coefficients $c_n$ to satisfy
\begin{equation}
 c_n=\delta_{n0}-\Delta t U G_0(n\Delta t).
\label{eq: cn_def}
\end{equation}
Since $G_0(t)$ is bounded, and decays exponentially fast in time, it is easy to see that $\sum_{n=-\infty}^\infty |c_n|<\infty$. We will assume for the moment that $\Ind C(\xi)=0$ (which turns out will be true at half filling when $U\lesssim 0.866$ on the hypercubic lattice, see below). Using the definition for $c_n$, we find
\begin{equation}
 C(e^{i\theta})=\sum_{n=-\infty}^\infty c_n e^{in\theta}=1-\Delta t U\sum_{n=-\infty}^\infty G_0(n\Delta t)e^{in\theta}.
\end{equation}
If we let $n\Delta t=t'$ and $\omega\Delta t=\theta$, we recognize that in the limit where $\Delta t\rightarrow 0$, the sum in the right hand side of the above equation approaches the Fourier transform of $G_0(t')$, so we write
\begin{equation}
 C(\omega)=1-U\int_{-\infty}^{\infty}dt' G_0(t')e^{i\omega t'}=1-UG_0(\omega),
\end{equation}
where $G_0(\omega)$ is defined by equation~(\ref{rGF_0_FT}). The coefficients $g_n$ in equation~(\ref{eq: g_t_def}) can then be written as
\begin{equation}
g(t')=\frac{g_n}{\Delta t}=\frac{1}{2\pi}\int_{-\pi/\Delta t}^{\pi/\Delta t}d\omega e^{-in\Delta t\omega}\ln C(e^{i\Delta t\omega})=\frac{1}{2\pi}\int_{-\infty}^{\infty}d\omega e^{-i\omega t'}\ln C(\omega),
\label{eq: g_t_def2}
\end{equation}
where we again used $t'=n\Delta t$.  Now, defining $D_N$ to satisfy
\begin{equation}
 D(t)=D(N\Delta t)=D_N=\det\nolimits_{[0,t]}\left\|\delta_{ij}-\Delta t G_{0}(t_i-t_j) U\right\|,
\end{equation}
we find from equation~(\ref{eq: szego}) that
\begin{equation}
 \lim_{t\rightarrow\infty}D(t)=\exp\left [ \frac{t}{2\pi}\int_{-\infty}^{\infty}d\omega \ln C(\omega)+\int_0^\infty dt' t' g(t')g(-t')\right ],
\label{eq: det_szego}
\end{equation}
which is the asymptotically exact result in the limit of large $t$ when the index of $C(\xi)$ is equal to zero. Since the coefficient of $t$ in the exponent can be a complex number, the determinant can oscillate while exponentially decaying at long times.  This occurs when the system has undergone the Mott transition and the $f$-electron spectral function displays a gap at low frequency.

But we can improve upon the result in equation~(\ref{eq: det_szego}) to provide finite-time corrections.  We illustrate next how to do this using the Wiener-Hopf sum equation approach. We start with our original $N+1$ simultaneous equations we need to solve in equation~(\ref{eq: wh1}), with the coefficients $c_n$ defined in equation~(\ref{eq: cn_def}). Next, we choose $y_n=\delta_{n0}$.  Cramers rule, applied to the first column of the $c$ matrix immediately tells us that
\begin{equation}
 x_0^{(N)}=\frac{D_N}{D_{N+1}},
\end{equation}
which further implies that
\begin{equation}
 D_N=\left ( \prod_{M=N}^\infty x_0^{(M)}\right ) \times D_{M\rightarrow\infty}.
\end{equation}
If we assume that $x_0^{(M)}=(1-\Delta t \delta_M)\exp(-g_0)$ (which we will verify below), then we find
\begin{equation}
 D(t)=\exp\left [ -\Delta t \sum_{M=N}^\infty \delta_M\right ] \exp\left [ \frac{t}{2\pi}\int_{-\infty}^{\infty}d\omega \ln C(\omega)+\int_0^\infty dt' t' g(t')g(-t')\right ].
\end{equation}
This will then provide the next order of corrections to the asymptotic limit of the determinant.

We need to find $x_0^{(M)}$ for large $M$.  Our strategy is to set $U(\xi)=0$ in equation~(\ref{eq: whfinal2}) to find $V(\xi^{-1})$, then solve for $U(\xi^{-1})$ with equation~(\ref{eq: whfinal4}) and the approximate $V(\xi)$, and finally substitute $U(\xi)$ into equation~(\ref{eq: whfinal1}) to find $X(\xi)$. The term $x_0^{(M)}$ then follows from taking the $\xi\rightarrow 0$ limit.  Using the facts that $Y(\xi)=1$ for our equation and $Q(\xi^{-1})-1$ is a $-$ function, we immediately find that equation~(\ref{eq: whfinal2}) can be solved by
\begin{equation}
 V(\xi^{-1})=-1+\left [ Q(\xi^{-1})\right ]^{-1}.
\end{equation}
Substituting into equation~(\ref{eq: whfinal4}) (after replacing $\xi\rightarrow\xi^{-1}$) then yields
\begin{equation}
 U(\xi^{-1})=-\left [ P(\xi^{-1})\right ]^{-1} \left [ P(\xi^{-1})\left \{ Q(\xi)\right \}^{-1}\xi^N\right ]_-.
\end{equation}
Now we must replace $\xi\rightarrow \xi^{-1}$ in the above equation.  This requires care because we will convert the $-$ function into a $+$ function, {\it but with no constant term in the power series expansion}. So we write
\begin{equation}
 U(\xi)=-\left [ P(\xi)\right ]^{-1} \left [ P(\xi)\left \{ Q(\xi^{-1})\right \}^{-1}\xi^{-N}\right ]_+^{\prime},
\end{equation}
with the prime indicating there is no constant term in the $+$ function.  This result can now be substituted into equation~(\ref{eq: whfinal1}) to yield
\begin{equation}
 X(\xi)=P(\xi)-P(\xi)\left [ Q(\xi^{-1})\left \{ P(\xi)\right \}^{-1}\xi^N\left [ P(\xi)\left \{ Q(\xi^{-1})\right \}^{-1}\xi^{-N}\right ]_+^{\prime}\right ]_+,
\label{eq: x_nowind}
\end{equation}
where we used the fact that $[Q(\xi^{-1})]_+=1$.  We will need to evaluate this in the limit $\xi\rightarrow 0$.

Note that we have the following two identities for $P$ and $Q$
\begin{equation}
 P(\xi)=\exp\left [ - \sum_{n=0}^\infty g_n\xi^n\right ]\quad {\rm and}\quad P(\omega)=\exp\left [
- \int_0^\infty dt g(t)e^{i\omega t}\right ],
\label{eq: p}
\end{equation}
and
\begin{equation}
 Q(\xi^{-1})=\exp\left [ - \sum_{n=-\infty}^{-1}g_n\xi^n\right ]\quad {\rm and}\quad Q(-\omega)=\exp\left [
- \int_{-\infty}^{-\Delta t} dt g(t)e^{i\omega t}\right ].
\label{eq: q}
\end{equation}
Using the definition for $g(t)$ in equation~(\ref{eq: g_t_def2}), one can immediately verify that $P(\omega)Q(-\omega)=1/C(\omega)$ in the limit as $\Delta t\rightarrow 0$, as it must.  Using the first relation in equation~(\ref{eq: p}), immediately shows us that $P(\xi=0)=\exp[-g_0]$. Evaluating the other term in equation~(\ref{eq: x_nowind}) requires more work. In order to create the relevant $+$ functions, we must first take the functions of $\omega$, Fourier transform them to functions of $t$ and then reverse Fourier transform back, but include only the positive time contributions.  For example, we can write
\begin{equation}
\left [ P(\xi)\left \{ Q(\xi^{-1})\right \}^{-1}\xi^{-N}\right ]_+^{\prime}=
\frac{1}{2\pi}\int_{\Delta t}^\infty dt' e^{i\omega t'}\int_{-\infty}^\infty d\omega' \frac{P(\omega')}{Q(-\omega')}e^{-i\omega' (t+t')}.
\end{equation}
The limit of $\Delta t$ is to remind us that there should be no constant term in the expansion.  Using this strategy, we can immediately write down the final answer for $x_0^{(N)}$.  We obtain $x_0^{(N)}=(1-\Delta t \delta_N)\exp[-g_0]$ with
\begin{equation}
 \delta_N=\frac{1}{4\pi^2}\int_{-\infty}^\infty d\omega \int_{\Delta t}^\infty dt'\int_{-\infty}^\infty d\omega' \frac{Q(-\omega)}{P(\omega)}\frac{P(\omega')}{Q(-\omega')}e^{i(\omega-\omega')(t+t')},
\end{equation}
where $t=N\Delta t$. Putting this all together finally yields the asymptotic expression for $D(t)$
\begin{equation}\label{eq:det finite time}
 D(t)=\exp\left [ \frac{t}{2\pi}\int_{-\infty}^{\infty}d\omega \ln C(\omega)+\int_0^\infty dt' t' g(t')g(-t')- \int_t^\infty d\bar t \int_{\Delta t}^\infty dt'h(t+t')h'(-t-t')\right ],
\end{equation}
with
\begin{equation}
 h(t)=\frac{1}{2\pi}\int_{-\infty}^\infty d\omega \frac{P(\omega)}{Q(-\omega)}e^{-i\omega t}\quad{\rm and}
\quad h'(t)=\frac{1}{2\pi}\int_{-\infty}^\infty d\omega \frac{Q(-\omega)}{P(\omega)}e^{-i\omega t}.
\end{equation}

\paragraph{Case 2: winding of --1.}

Szeg\"o's theorem fails when the index of $C(\xi)$ is nonzero, and $\ln C(\xi)$ winds around the origin.  This occurs on the hypercubic lattice at half filling when $U\gtrsim0.866$, see below.  We can still derive asymptotic formulas for the Green's function by solving an auxiliary problem, which has the winding removed.  We now show how this is done.

In all numerical cases we have examined, the index of $C(\xi)$ satisfies $\Ind C(\xi)=-1$ when $U$ is large enough.  Such a winding can  be removed by considering $\bar C(\xi)=-\xi C(\xi)$ which has index zero (the minus sign is introduced for convenience, as will be clear below).  The coefficients obviously satisfy
\begin{equation}
 \bar c_n=-c_{n-1}.
\end{equation}
We examine the auxiliary Toeplitz determinant
\begin{displaymath}
 \bar D_N=\left | 
\begin{array}{ccccc}
 \bar c_0 & \bar c_{-1} & \bar c_{-2}&\hdots & \bar c_{-N+1}\\
\bar c_1 & \bar c_0 & \bar c_{-1}&\hdots &\bar c_{-N+2}\\
\bar c_2 & \bar c_1 & \bar c_0 & \hdots &\bar c_{-N+3}\\
\vdots & \vdots & \vdots & \vdots & \vdots\\
\bar c_{N-1} & \bar c_{N-2} & \bar c_{N-3}&\hdots & \bar c_0
\end{array}
\right | = \left |
\begin{array}{ccccc}
 -c_{-1} & -c_{-2} & -c_{-3}&\hdots & -c_{-N}\\
-c_0 & -c_{-1} & -c_{-2}&\hdots & -c_{-N+1}\\
 -c_1 &  -c_0 &  -c_{-1} & \hdots & -c_{-N+2}\\
\vdots & \vdots & \vdots & \vdots & \vdots\\
 -c_{N-2} &  -c_{N-3} &  -c_{N-4}&\hdots &  -c_{-1}
\end{array}
\right |,
\end{displaymath}
and consider the $N+1$ simultaneous equations in equation~(\ref{eq: wh1}) with the barred variables
\begin{equation}
 \sum_{m=0}^N\bar c_{n-m}\bar x^{(N)}_m=\bar y_n
\end{equation}
with $\bar y_n=\delta_{n0}$.  Then Cramers rule, evaluated for the $N+1$st column, tells us
\begin{equation}
 \bar x_N^{(N)}=D_{N-1}/\bar D_N,
\end{equation}
where the two factors of $(-1)^N$ cancel.
Szeg\"o's theorem can be applied to $\bar C(\xi)$, so we immediately learn that
\begin{equation}
 \lim_{N\rightarrow\infty}\bar D_N=\exp\left [ N \bar g_0 + \sum_{n=1}^\infty n \bar g_n \bar g_{-n}\right ],
\label{eq: szego_bar}
\end{equation}
with
\begin{equation}
 \bar g_n=\frac{1}{2\pi}\int_{-\pi}^{\pi}d\theta e^{-in\theta}\ln \bar C(e^{i\theta}).
\label{eq: g_t_def_bar}
\end{equation}
We need to calculate $\bar x_N^{(N)}$ to find the determinant $D_N$.

First note that equations~(\ref{eq: whfinal1}--\ref{eq: whfinal4}) hold for the barred functions when we have a winding around the origin. We will once again solve for $\bar V(\xi^{-1})$ setting $\bar U(\xi)=0$, then substitute into equation~(\ref{eq: whfinal3}) to find $\bar X(\xi^{-1})\xi^N$. Taking the limit $\xi\rightarrow 0$ will then give us $\bar x_N^{(N)}$. Just like before, we find
\begin{equation}
 \bar V(\xi)=-1+\left [ \bar Q(\xi)\right ]^{-1},
\end{equation}
which then gives
\begin{equation}
 \bar X(\xi^{-1})\xi^N=\bar Q(\xi) \left [ \left \{ \bar Q(\xi)\right \}^{-1}\bar P(\xi^{-1})\xi^N\right ]_+.
\end{equation}
We take the limit $\xi\rightarrow 0$ by integrating the function on the right hand side over the unit circle and dividing by $2\pi$.  Using $\theta=\Delta t \omega$, then yields
\begin{equation}
 \bar x_N^{(N)}=\frac{\Delta t}{2\pi}\int_{-\pi/\Delta t}^{\pi/\Delta t} d\omega  e^{i\omega t}\bar P(-\omega)/\bar Q(\omega),
\label{eq: x_bar_final}
\end{equation}
where we used $t=N\Delta t$ again.  Note that the limits on the integration cannot be simply extended to $\pm\infty$ as we did before.  Since the function $\ln C(\exp[i\theta])$ winds once around the origin, it displays a discontinuity of $-2i\pi$ to its imaginary part as $\theta$ runs from $-\pi$ to $\pi$. We remove this discontinuity by adding a linear function that is equal to $0$ at $\theta=-\pi$ and is equal to $2i\pi$ at $\theta=\pi$ (the minus sign in $\bar C$ is needed to move the branch cut of the logarithm from the positive real axis to the negative real axis).  Since this linear function cannot be extended to infinity, we need to work with a finite value of $\Delta t$ in the calculations
(for a further discussion, look at Fig.~\ref{fig: ln_c} and the corresponding discussion below).  We simply take $\Delta t$ small enough, that we see the numerical results do not change, and the system has approached its $\Delta t\rightarrow 0$ limit.
The final result for the determinant is then
\begin{equation}\label{eq: det winding}
 D(t)=\exp\left [ \frac{t}{2\pi} \int_{-\infty}^{\infty}d\omega \ln \bar C(\omega) + \int_0^\infty dt't'\bar g(t')\bar g(-t')\right ]\frac{\Delta t}{2\pi}\int_{-\pi/\Delta t}^{\pi/\Delta t} d\omega'  e^{i\omega' t}\bar P(-\omega')/\bar Q(\omega').
\end{equation}
The limits on the first integral have been extended to $\pm\infty$, because it turns out that the linear piece gives no contribution to the integral.  The definitions of $\bar g(t)$, $\bar P(\omega)$ and $\bar Q(\omega)$ are obvious from the above discussion.

It should be noted that, in contrast to the case without winding where the third term in the exponent of equation~(\ref{eq:det finite time}) gives finite time corrections to the large $t$ exponential asymptotics of equation~(\ref{eq: det_szego}), in the case with winding, equation~(\ref{eq: det winding}) has a time dependent prefactor with a nontrivial $\Delta t\to 0$ and $t\to\infty$ limit which can not be considered as a vanishing correction to the exponential decay of equation~(\ref{eq: det_szego}).

\section{Numerical results}

\begin{figure}[htb]
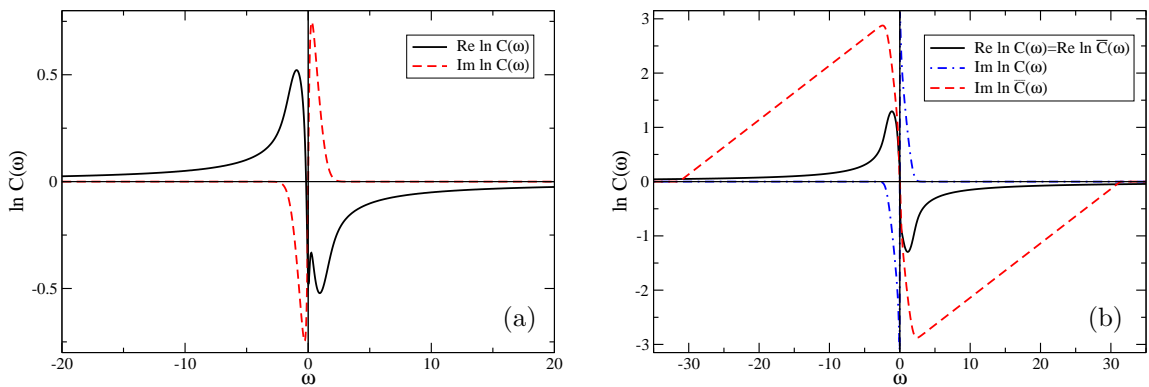

 \centering
 \includegraphics[scale=0.3,clip]{windU05T01}%
\unitlength=1mm%
\begin{picture}(0,0)
 \put(-8,8){\makebox{(a)}}
\end{picture}%
\hfill
 \includegraphics[scale=0.3,clip]{windU15T01}%
\unitlength=1mm%
\begin{picture}(0,0)
 \put(-8,8){\makebox{(b)}}
\end{picture}%
 \caption{(Color online) Plot of (a) $\ln C(\omega)$ for the case with no winding [$U=0.5$, $\Ind C(\xi)=0$] and (b) $\ln C(\omega)$ and $\ln \bar C(\omega)$ for the case with winding [$U=1.5$, $\Delta t=0.1$, $\Ind C(\xi)=-1$]. }
\label{fig: ln_c}
\end{figure}

We begin our numerical discussion by describing the different behavior of $C(\xi)$ 
when it has no winding or when it winds with an index of $-1$ around the origin.  In Fig.~\ref{fig: ln_c}, we show a plot of $\ln C(\xi)$ for two cases: (a) the case with no winding, and (b) the case with an index of $-1$.  These two cases are distinguished by the sign of the real part of $C(\xi)$ at $\xi=0$, where its imaginary part changes sign. On the hypercubic lattice at half filling we have $\Real C(0)>0$ for $U<0.866$ and there is no winding and $\Real C(0)<0$ for $U>0.866$ and now the index is equal to $-1$.  Notice how the imaginary part has a steep change in its value near $\xi=0$ [panel (a)], which evolves into a discrete jump by $2\pi i$ when $U\gtrsim 0.866$ [panel (b)], the jump at the origin is compensated by the linear shift which runs from the minimal to maximal values of the frequency used in the calculations ($|\omega|=10\pi$ here). While it is obvious that integration over $\omega$ can easily be extended to $\omega=\pm \infty$ when there is no winding, one needs to carefully include contributions present at the given value of $\Delta t$ when there is winding, and one cannot extend the integration limits in this case.

\begin{figure}[htb]
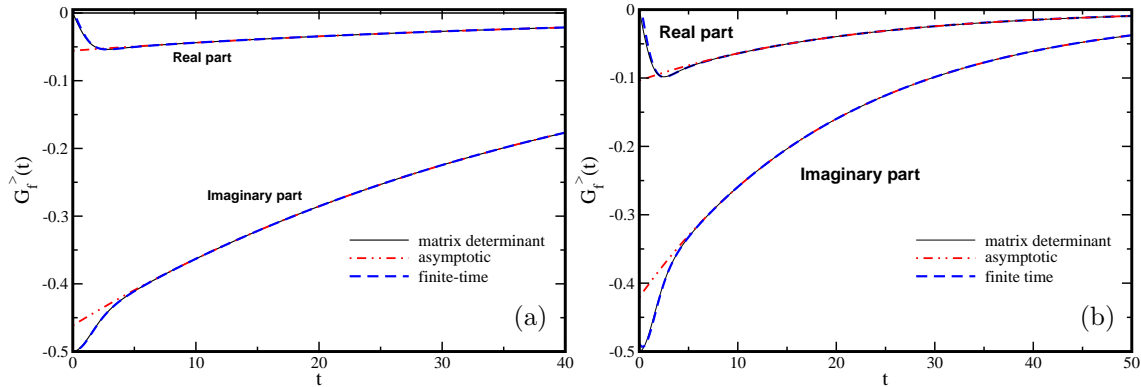

 \centering
 \includegraphics[scale=0.3,clip]{U05T01}%
\unitlength=1mm%
\begin{picture}(0,0)
 \put(-8,8){\makebox{(a)}}
\end{picture}%
%\hfill
 \includegraphics[scale=0.3,clip]{U07T01}%
\unitlength=1mm%
\begin{picture}(0,0)
 \put(-8,8){\makebox{(b)}}
\end{picture}%
 \caption{(Color online) Real and imaginary parts of the greater Green's function for cases with no winding (a) $U=0.5$ and (b) $U=0.7$ ($T=0.1$). 
The solid (black) lines represent exact results of the functional determinant in equation~(\ref{eq: g_greater_discrete}),
the dot-dashed (red) line represents the asymptotic result in equation~(\ref{eq: det_szego}), and the dashed (blue) line represents the expression with finite-time corrections in equation~(\ref{eq:det finite time}). Note that we cannot really discern any difference between the exact matrix results and the asymptotic Wiener-Hopf results all the way down to $t=0$. }
\label{fig: nowinding}
\end{figure}

Next we compare our different asymptotic expansions to the results derived from a direct numerical calculation with the matrix formalism from equation~(\ref{eq: g_greater_discrete}) when $T=0.1$.  This involves a straightforward approach where we use three different discretization sizes ($\Delta t=0.1$, 0.0666, and 0.0333) which we extrapolate to $\Delta t\rightarrow 0$, and we compare those results to calculations with different approximations.  In the case with no winding, we use both the asymptotic form for large $t$ in equation~(\ref{eq: det_szego}) and the more correct form for smaller $t$ in equation~(\ref{eq:det finite time}).  In Fig.~\ref{fig: nowinding}, we show the results for the greater Green's function at half filling on the hypercubic lattice for $U=0.5$ [panel (a)] and $U=0.7$ [panel (b)].   Note how the exponential form, that comes from Szeg\"o's theorem, is quite accurate for large times, but becomes poor for small times, and how the corrections arising from the Wiener-Hopf approach produce remarkable agreement with the numerically exact results for essentially all $t$.  The errors are the largest near $t=0$, but even there, they lie below the few percent level.  This shows that the analytic formulas are quite accurate for all $t$ when there is no winding.

\begin{figure}[htb]
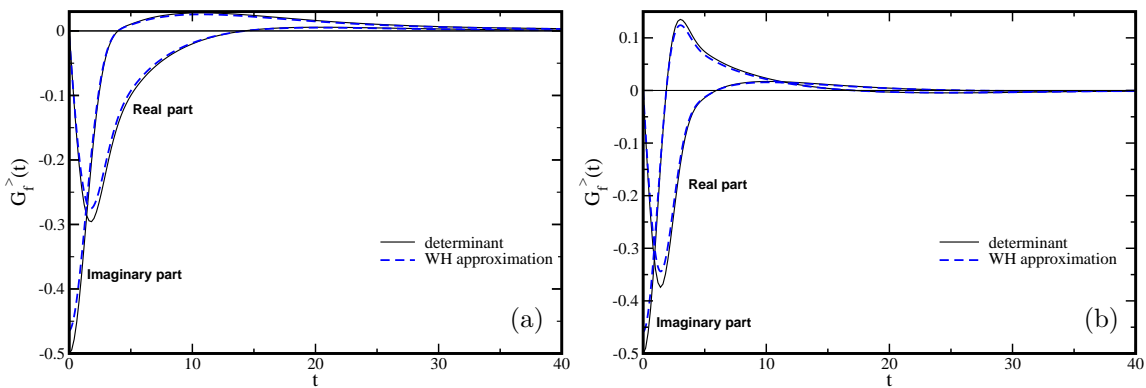

 \includegraphics[scale=0.3,clip]{U15T01}%
\unitlength=1mm%
\begin{picture}(0,0)
 \put(-8,8){\makebox{(a)}}
\end{picture}%
\hfill
 \includegraphics[scale=0.3,clip]{U20T01}%
\unitlength=1mm%
\begin{picture}(0,0)
 \put(-8,8){\makebox{(b)}}
\end{picture}%
 \caption{(Color online) Real and imaginary parts of the greater Green's function for the case with winding [$\Ind C(\xi)=-1$] (a) $U=1.5$ and (b) $U=2$ ($T=0.1$). 
The solid lines represent the extrapolated matrix results for the functional determinant in equation~(\ref{eq: g_greater_discrete}) and the
dashed line represents the asymptotic expression with the finite-time corrections in equation~(\ref{eq: det winding}). }
\label{fig: winding}
\end{figure}

Next we examine what happens in the case with winding.  We examine two values of $U$ in Fig.~\ref{fig: winding}: $U=1.5$ which is near the critical interaction for the metal-insulator transition $U_c=\sqrt{2}$, and $U=2$ which is a small gap insulator. We show the results from the scaled matrix calculations with a solid line (black) and the asymptotic Wiener-Hopf results with a (blue) dashed line. In this case, the Green's function has behavior that does not approach the asymptotic exponential behavior for short times, and the analytic formula is less accurate for small times (with maximal error is on the order of 10\%).

Now we can specify the differences between the cases without winding and with winding. For $U<0.866$ there is no winding and our exact results of the functional determinants rapidly approach the asymptotic result (with finite-time corrections) in equation~(\ref{eq:det finite time}) and we do not observe a crossing of the zero axis at any temperature and at $T\to0$ the long-time behavior is replaced by a
power law. On the other hand, for $0.866<U<U_c$ we always observe a crossing of the axis at high temperatures, which originates from the time dependent prefactor in equation~(\ref{eq: det winding}). With decreasing temperature, the crossing point shifts to larger times producing some kind of exponential decay for intermediate values of $t$ and at $T=0$ we observe a crossover to power law behavior when the zero crossing is pushed out to infinity. For $U>U_c$ the crossing of the axis is observed at finite time values for any temperature and does not transform into power law at zero temperature.

The Fourier transforms to the spectral formula produces results essentially equivalent to those shown already in Refs.~\cite{freericks_turkowski_zlatic,freericks_turkowski_zlatic2,freericks_turkowski_zlatic3}, so we do not repeat them here.

\section{Conclusions}

In this work, we have shown a new representation for the $f$-electron Green's function, which allows for efficient numerical computation.  By using the property that $U_c(t,t')$ is nonzero only on the interval $[0,t]$, we restrict the determinant of the continuous matrix operators to a finite time interval instead of over the three branches of the Kadanoff-Baym-Keldysh contour. This produces a huge savings in the computational effort, because the size of the matrices do not grow with temperature and they are less than one half the size of the matrices used in previous numerical calculations. As a result, we can examine much wider regions of parameter space.

In addition, we used the Wiener-Hopf sum equation approach, along with Szeg\"o's theorem, to derive exact analytical expressions for the Toeplitz determinants required to find the Green's functions. While these expressions are asymptotically exact in the limit where $t\rightarrow\infty$, we find they have errors less than 10\% all the way down to $t=0$. This then allows us to use the matrix results for small times, and then append them by the asymptotic expressions for large times in order to find accurate results for the Green's function at all times.  These are then Fourier transformed to real frequencies to determine the $f$-electron spectral function. 

The $f$-electron spectral function displays the expected behavior as well.  For $U$ values smaller than the critical $U$ for the Mott transition ($U=\sqrt{2}$), the spectral function develops a sharp peak at low $T$, which ultimately diverges as an inverse power law in $\omega$ as $T\rightarrow 0$ [rigorously speaking, our asymptotic formulas are not valid at $T=0$, because the function $\ln C(\xi)$ becomes discontinuous due to the jump in the Fermi factor, but the power law behavior can be extracted via other techniques---our focus here was on the finite-$T$ behavior]. When $U$ is larger than the critical $U$, a gap forms in the $f$-electron spectral function (rigorously speaking, it is a pseudogap on the hypercubic lattice) but subgap states rapidly enter as a function of $T$ for low $T$.

\section*{Acknowledgments}

It is with great pleasure that we honor Prof. Stasyuk with this contribution. This publication is based on work supported by Award No. UKP2-2697-LV-06 of the U.S. Civilian Research \& Development Foundation (CRDF). J.K.F. also acknowledges support by the National Science Foundation under grant number DMR-0705266.

\label{last@page}

\begin{thebibliography}{99}
\bibitem{falicov_kimball} Falicov~L.M., Kimball~J.C., Phys. Rev. Lett., 1969, \textbf{22}, 997.
\bibitem{freericks_review} Freericks~J.K., Zlati\'c~V., Rev. Mod. Phys., 2003, \textbf{75}, 1333.
\bibitem{stasyuk:FKM1} Stasyuk~I.V., Shvaika~A.M., J. Phys. Studies, 1999, \textbf{3}, 177.
\bibitem{stasyuk:FKM_f1} Stasyuk~I.V., Hera~O.B., Condens. Matter Phys., 2003, \textbf{6}, 127.
\bibitem{stasyuk:FKM_f2} Stasyuk~I.V., Hera~O.B., Phys. Rev.~B, 2005, \textbf{72}, 045134.
\bibitem{stasyuk:FKM_f3} Stasyuk~I.V., Hera~O.B., Eur. Phys. J. B, 2005, \textbf{48}, 339.
\bibitem{brandt_urbanek} Brandt~U., Urbanek~M.P., Z. Phys. B: Condens. Matter, 1992, \textbf{89}, 297.
\bibitem{janis} Janis~V., Phys. Rev. B, 1994, \textbf{49}, 1612;  Physica B, 1996, \textbf{223--224}, 616;  Int. J. Mod. Phys. B, 1997 \textbf{11}, 433.
\bibitem{zlatic_review} Zlati\'c~V., Freericks~J.K., Lema\'nski~R., Czycholl~G, Philos. Mag. B, 2001, \textbf{81}, 1443.
\bibitem{freericks_turkowski_zlatic} Freericks~J.K., Turkowksi~V.M., Zlati\'c~V., Phys. Rev. B, 2005, \textbf{71}, 115111.
\bibitem{freericks_turkowski_zlatic2} Freericks~J.K., Turkowksi~V.M., Zlati\'c~V., in  {\it Proceedings of the HPCMP Users Group Conference 2004, Williamsburg, VA, June 7--11, 2004} edited by R. E. Peterkin, IEEE Computer Society, Los Alamitos, CA, 2004, p. 7.
\bibitem{freericks_turkowski_zlatic3} Freericks~J.K., Turkowksi~V.M., Zlati\'c~V., Physica B, 2005 \textbf{359--361}, 684.
\bibitem{anders_czycholl}  Anders~F.B., Czycholl~G., Phys. Rev. B, 2005, \textbf{71}, 125101.
\bibitem{liu} Liu~Y.-L., Phys. Rev. B, 2005, \textbf{ 72}, 045123.
\bibitem{rutgers} M\"oller~G., Ruckenstein~A.E., Schmitt-Rink~S., Phys. Rev. B, 1992, \textbf{46}, 7427.
\bibitem{mahan} Mahan~G.D., Phys. Rev., 1967, \textbf{163}, 612.
\bibitem{nozieres_dedominici} Nozi\`eres~P., De Dominicis~C., Phys. Rev., 1969, \textbf{178}, 1097.
\bibitem{kadanoff_baym} Kadanoff~L.P., Baym~G., {\it Quantum statistical mechanics}, Benjamin, New York, 1962.
\bibitem{keldysh} Keldysh~L.V., Zh. Eksp. Teor. Fiz., 1964, \textbf{47}, 1945 (in Russian) [Sov. Phys. JETP, 1965, \textbf{20}, 1018].
\bibitem{mccoy_wu} McCoy~B.M.,Wu~T.T., {\it The two-dimensional Ising model}, Harvard University Press, Cambridge, MA, 1973.
\bibitem{wiener_hopf} Wiener~N., Hopf~E., Sitzber. Preuss. Akad. Wiss. Berlin, Kl. Math. Phys. Tech., 1931, 696.
\bibitem{krein} Krein~M.G., Uspekhi Mat. Nauk, 1958, \textbf{13}, 3 (in Russian) [Am. Math. Soc. Transl., 1962, \textbf{22}, 163].
\bibitem{szego} Szeg\"o~G., Math. Z., 1920, \textbf{6}, 167; Math. Z., 1921, \textbf{9}, 167.
\bibitem{szego2} Szeg\"o~G., Commun. Sem. Math. Univ. Lund, suppl. ded. Marcel Reisz, 1952, 228.
\bibitem{szego3} Grenander~U., Szeg\"o~G., {\it Toeplitz forms and their applications}, University of California Press, Berkeley and Los Angeles, 1958.

%     \bibitem{Zub74} Zubarev~D.N., Nonequilibrium Statical Thermodynamics.
%         Consultant Bureau, New-York, 1974.
%     \bibitem{Bus89} Bussman-Holder~A., Simon~A., Buttner~H.,
%         Phys. Rev.~B, 1989, \textbf{39}, 207.
%     \bibitem{Mel00} Melnyk~R.S., Patsahan~O.V.,
%         Teor. Mat. Fiz., 1987, \textbf{124}, No.~2, 339 (in Russian)
%         [Theor. Math. Phys., \textbf{124}, 1145].
%     \bibitem{Mry94} Mryglod~I.M., Tokarchuk~M.V.,
%         Condens. Matter Phys., 1994, \textbf{3}, 116.
%     \bibitem{Sta04} Stasyuk~I.V., Mysakovych~T.S.
%         Preprint of the Institute for Condensed Matter Physics, ICMP--04--12U,
%         Lviv, 2004 (in Ukrainian).
%     \bibitem{Shv04} Shvaika~A.M., Vorobyov~O., Freericks~J.K., Devereaux~T.P.
%         Preprint arXiv:cond-mat/0408400,
%         2004.
%     \bibitem{Yuk87} Yukhnovkii~I.R., Idzyk~I.M., Kolomiets~V.O.
%         Critical point of the liquid-gas system. -- In: Proc. Contributed
%         papers of Conf. on Modern problems of statistical physics. Vol.~2,
%         Lviv, 3--5 February 1987, p.~97--102 (in Russian).
\end{thebibliography}
\end{document}